\documentclass[aps,showpacs,amsmath,amssymb,pra,floatfix,superscriptaddress,a4,twocolumn]{revtex4}

\usepackage[dvips]{graphicx}
\usepackage{amsmath}
\usepackage{color}

\def\CR{\nonumber\\[0.15cm]}
\newcommand{\ket}[1]{|\,{#1}\,\rangle}
\newcommand{\braket}[2]{\mbox{$\langle\,{#1}\, | \,{#2}\,\rangle$}}
\newcommand{\pdiff}[2]{\frac{\partial #1}{\partial #2}}

\newcommand{\eref}[1]{Eq.~(\ref{#1})}

\newcommand{\aref}[1]{Appendix \ref{#1}}
\def\CR{\nonumber\\[0.15cm]}
\newcommand{\sub}[2]{{#1}_{\mbox{\!\! \scriptsize #2}}}
\newcommand{\bv}[1]{\mathbf{ #1 }}

\definecolor{orange}{rgb}{0.8,0.4,0.0} 
\newcommand{\sew}[1]{\textcolor{black}{#1}}

\newcommand{\mge}[1]{\textcolor{black}{#1}}

\begin{document}

\title{Break-up of Rydberg superatoms via dipole-dipole interactions}

\author{S. M{\"o}bius}
\affiliation{Max-Planck-Institute for the Physics of Complex Systems, 01187 Dresden, Germany}

\author{M. Genkin}
\affiliation{Max-Planck-Institute for the Physics of Complex Systems, 01187 Dresden, Germany}

\author{S. W{\"u}ster}
\affiliation{Max-Planck-Institute for the Physics of Complex Systems, 01187 Dresden, Germany}

\author{A. Eisfeld}
\affiliation{Max-Planck-Institute for the Physics of Complex Systems, 01187 Dresden, Germany}

\author{J. M. Rost}
\affiliation{Max-Planck-Institute for the Physics of Complex Systems, 01187 Dresden, Germany}

\date{\today}

\begin{abstract}
We investigate resonant dipole-dipole interactions between two ``superatoms'' of different angular momentum, consisting of two Rydberg-blockaded atom clouds
where each of them carries initially a coherently shared single excitation. 
We demonstrate that the dipole-dipole interaction breaks up the superatoms by removing the excitations from the clouds. The dynamics is 
akin to an ensemble average over systems where only one atom per cloud participates in entangled motion and excitation transfer. 
Our findings should thus facilitate the experimental realization of adiabatic exciton transport in Rydberg systems by replacing single sites with atom clouds.

\end{abstract}


\maketitle

%%%%%%%%%%%%%%%%%%%%%%%%%%%%%%%%%%%%%%
\section{Introduction}
%%%%%%%%%%%%%%%%%%%%%%%%%%%%%%%%%%%%%%

Rydberg atoms have several remarkable 
properties, such as long lifetimes, large polarizability and strong long-range interactions. These properties have  turned Rydberg atoms into versatile tools for
quantum information~\cite{Jaks_00,Safr_03,Saff_09,Saff_10,Isen_10}, nonlinear quantum optics~\cite{Weat_08,Prit_10,Ates_11} or 
cavity quantum electrodynamics~\cite{Yama_02,Yang_07,Guer_10}. Recently, also the potential of Rydberg atoms for 
quantum transport in atomic aggregates has been demonstrated~\cite{West_06,Mulk_07,Ates_08,Wust_10,Moeb_11}.
The underlying physical mechanism is closely related to the one of excitation transport in molecular systems~\cite{Mulk_07,Holstein:model,Eisf_02}.

So far, only single atoms have been considered as building blocks for excitation transport in Rydberg aggregates. 
An experimental implementation would be facilitated if the single atom sites could be replaced by atom clouds. This has motivated us to explore the possibilities for extending the existing Rydberg transport schemes 
using atom clouds as sites.

When several atoms are brought together in clouds, they exhibit an effect known
as Rydberg blockade~\cite{Luki_01,Tong_04,Sing_04,Robi_05,Vogt_07,Heid_07,atpo+07,Urba_09,Gaet_09,Vite_11}. At small interatomic distances, the
van-der-Waals interactions $U_{\rm vdW}=C_6/r^6$ (where $C_6$ is a state-dependent interaction constant and $r$ the interatomic distance) become large.
These interactions lead to an energy offset of all many-particle states with more than one Rydberg excitation with respect to the energy of the states with just a single excitation.
As a consequence, although irradiating an entire atomic cloud, a  laser can create at most one Rydberg excitation resonantly, within a radius at which this energy offset is larger than the laser linewidth. The latter 
is limited by the Rabi frequency $\Omega$ of the driven transition, which gives an estimate of the minimum interatomic distance 
between two Rydberg excitations, known as the blockade radius $r_{\rm bl}\approx (C_6/\hbar\Omega)^{1/6}$.
The interplay of the Rydberg blockade with quantum transport has appealing new aspects which we investigate in the following.

We concentrate on a system of two atom clouds, well separated in space. A similar arrangement was studied in Ref.~\cite{Carr_09} in the limit of a frozen Rydberg gas.
Here we choose the spatial extension $\sigma$ of each cloud to be smaller than the blockade radius $r_{\rm bl}$,  
while the inter-cloud distance $L$ is significantly larger than $r_{\rm bl}$, see Fig.~\ref{fig0}. 
For such parameters, the electron dynamics is restricted to a single Rydberg excitation per cloud, and we
consider a situation in which each of the two Rydberg-blockaded  
clouds is initially prepared in a coherent collective excitation. Such states can nowadays be experimentally created~\cite{Heid_07} and are
sometimes referred to as superatoms~\cite{Robichaux:superatoms,vuletic:superatoms}. They are a coherent superposition of states where all atoms within the cloud but one are in the ground state and one atom is in a Rydberg 
state $|\nu l\rangle$.
Here, $\nu$ and $l$ denote the principal and angular momentum quantum numbers of the Rydberg state, respectively.
We shall further imply with the term superatom, that all the atoms taking part in the superposition of electronic states share almost identical spatial probability distributions. 

We will analyze the resonant dipole-dipole interactions~\cite{Ande_98,Robi_04,Li_05} between two superatoms of different angular momentum,
with the setup and the model described in section~\ref{sec2}. Then, we
will  discuss the electronic excitation transfer in the limit of frozen atomic positions in section~\ref{sec3}, which already singles out some interesting features of the two interacting superatoms.
Subsequently, we will turn to the effects of the electronic excitations on the motion of the atoms in section~\ref{sec4}. 
The paper ends with a summary and conclusion in section~\ref{sec6}. Relevant derivations regarding the semisclassical nature of electronic dephasing,
the separability of free and entangled atomic motion, the numerical implementation of the atomic dynamics and collisional influences are provided in Appendices A-D.

%%%%%%%%%%%%%%%%%%%%%%%%%%%%%%%%%%%%%%
\section{The setup}\label{sec2}
%%%%%%%%%%%%%%%%%%%%%%%%%%%%%%%%%%%%%%

\begin{figure}
\begin{footnotesize}
\begin{center}
\scalebox{0.2}{\includegraphics{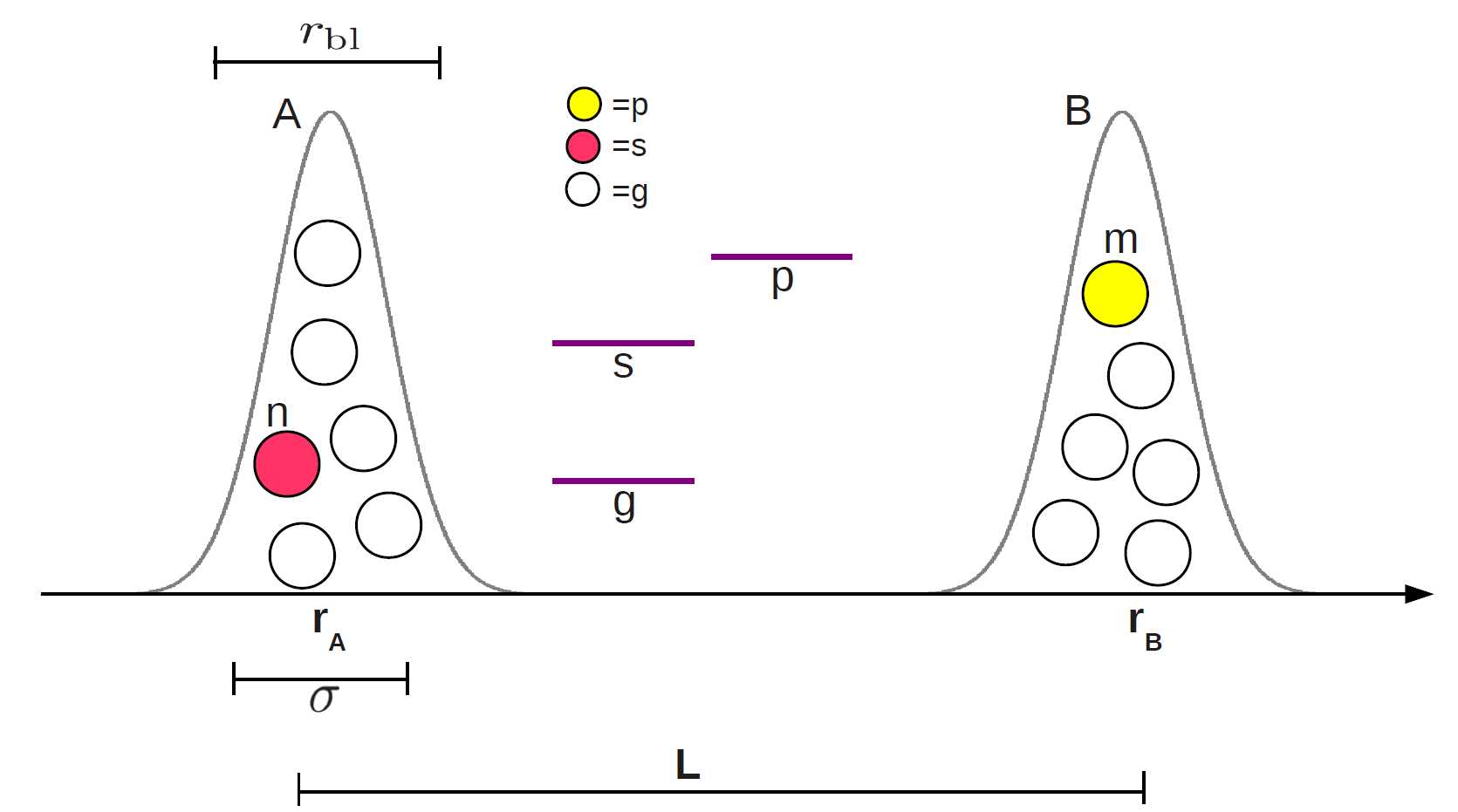}}
\end{center}  
\begin{quote} 
\begin{center}
\caption{Color online: Sketch of two interacting Rydberg blockaded clouds \mge{(not drawn to scale)}. The width $\sigma$ of the Gaussian distribution of the atoms
within each cloud is smaller than the blockade radius $r_{\rm bl}$, while the distance $L$ between the clouds is larger. The coloring (shading)
illustrates the electronic basis states $|\pi_{nm}\rangle$ introduced in Eq.~(\ref{elbasis}).}
\label{fig0}
\end{center}  
\end{quote}
\end{footnotesize}
\end{figure}

We denote the two atom clouds as $A$ and $B$ with  $N_A$  
and $N_B$ atoms in the respective cloud, while  $N=N_A+N_B$ is the total number of atoms. 
Each atom is labeled uniquely, such that atoms with numbers $1..N_A$ belong to cloud $A$ while $(N_A+1)..N$ refers to atoms in cloud $B$, summarized in the index sets
\begin{equation}
\mathcal{A}=[1,..,N_A],\quad \mathcal{B}=[N_A+1,..,N].
\end{equation}
We consider one spatial dimension along the separation of the clouds, and take three electronic states into account:
The ground state $|g\rangle$ and two Rydberg states $|\nu s\rangle$ and $|\nu p\rangle$. Since the principal quantum number $\nu$ of the Rydberg states is kept fixed
it will be omitted in the following. 
With these restricitions, the Hamiltonian for our system reads
\begin{equation}\label{fullHam}
H(r_1,...,r_N)=-\sum_{i=1}^N\frac{\hbar^2}{2M}\nabla^2_{r_i}+H_{\rm el}(r_1,...,r_N),
\end{equation}
where $M$ is the mass of a single atom, $r_i$ the position of the $i$-th atom and $H_{\rm el}$ the electronic Hamiltonian describing dipole-dipole-interactions between the atoms.
Generally, any of the $N$ atoms can be in either of the three states $|g\rangle$, $|s\rangle$, $|p\rangle$. 
However, the Rydberg blockade and the binary character of the dipole-dipole interactions significantly reduce the dimensionality of our problem. 
The former excludes states with more than one Rydberg excitation with {\it the same} angular quantum number within one cloud, while the latter excludes states in which both the s {\it and} the p 
excitation are localized within one cloud.
We define electronic states
\begin{equation}\label{elbasis}
|\pi_{nm}\rangle=|ggg..s..ggg..p..ggg\rangle,
\end{equation}
where the $n$-th atom is in the Rydberg state $|s\rangle$, the $m$-th atom in the Rydberg state $|p\rangle$ and all others in the ground state, and impose an additional constraint
\begin{equation}\label{eq:sets}
(n\in\,\mathcal{A}\,\,{\rm and}\,\,m\in\,\mathcal{B})\,\,{\rm OR}\,\,(m\in\,\mathcal{A}\,\,{\rm and}\,\,n\in\,\mathcal{B})\,,
\end{equation}
as sketched in Fig.~\ref{fig0}. 
The states of Eq.~(\ref{elbasis}) constrained according to Eq.~(\ref{eq:sets}) thus constitute a basis in the $2N_AN_B$-dimensional electronic subspace. 
We assume that 
the dipole-dipole interaction of strength $V_{0}$
\begin{equation}\label{dipdip}
V_{nm}(r_{nm})=\frac{V_0}{r_{nm}^3}
\end{equation}  
between the two excited
atoms $n$ and $m$ depends only on the interatomic distance $r_{nm}=|r_n-r_m|$ neglecting  
possible orientation effects. This is a description sufficient for certain geometries and selected states~\cite{Carr_09,Moeb_11,park:dipolebroadening}.
Using Eq.~(\ref{dipdip})  the matrix elements of the electronic Hamiltonian $H_{\rm el}$ from Eq.~(\ref{fullHam}) read
\begin{equation}\label{Hel}
\langle\pi_{nm}|\,H_{\rm el}\,|\pi_{n'm'}\rangle=V_{nm}\delta_{nm'}\delta_{mn'}.
\end{equation}
\begin{table}[h]
\caption{Matrix elements of the electronic Hamiltonian $H_{\rm el}$ in the basis $|\pi_{nm}\rangle$ for the example of $N_A=N_B=2$.}
\label{tab1}
\begin{center}
\begin{tabular}[c]{c|cc|cc|cc|cc}
$H_{\rm el}$  & $|\pi_{13}\rangle$ & $|\pi_{31}\rangle$ & $|\pi_{14}\rangle$ & $|\pi_{41}\rangle$ & $|\pi_{23}\rangle$ & $|\pi_{32}\rangle$ & $|\pi_{24}\rangle$ & $|\pi_{42}\rangle$   \\
\hline
$\langle\pi_{13}|$ & 0 & $V_{13}$ & 0 & 0 & 0 & 0 & 0 & 0 \\
$\langle\pi_{31}|$ & $V_{13}$ & 0 & 0 & 0 & 0 & 0 & 0 & 0 \\
\hline
$\langle\pi_{14}|$ & 0 & 0 & 0 & $V_{14}$ & 0 & 0 & 0 & 0 \\       
$\langle\pi_{41}|$ & 0 & 0 & $V_{14}$ & 0 & 0 & 0 & 0 & 0 \\
\hline
$\langle\pi_{23}|$ & 0 & 0 & 0 & 0 & 0 & $V_{23}$ & 0 & 0 \\
$\langle\pi_{32}|$ & 0 & 0 & 0 & 0 & $V_{23}$ &0 &  0 & 0 \\
\hline
$\langle\pi_{24}|$ & 0 & 0 & 0 & 0 & 0 & 0 & 0 & $V_{24}$ \\
$\langle\pi_{42}|$ & 0 & 0 & 0 & 0 & 0 & 0 & $V_{24}$ & 0 \\
%\hline
%\hline
\end{tabular}
\end{center}
\end{table} 
As a consequence of the Rydberg blockade, the electronic Hamiltonian in the matrix representation, Eq.~(\ref{Hel}), has a block-diagonal structure,
as can be seen in Table~\ref{tab1}. This has drastic consequences for the electron and atom dynamics: (i) The excitation exchange between the clouds
can be resembled by an average over single atom pairs and  (ii) the force resulting from the dipole-dipole interactions only acts on one atom per cloud rather than the whole cloud. 
We will discuss these effects in sections \ref{sec3} and \ref{sec4}, respectively. Before doing so, we introduce a few necessary technical details in the following subsection.

\subsection{Wave function respresentation}
The full wave function can be written as an expansion in the electronic basis $|\pi_{nm}\rangle$
\begin{equation}\label{fullwf}
\ket{\Psi({\bf R},t)}=\sum_{nm}\tilde{\phi}_{nm}({\bf R},t)|\pi_{nm}\rangle
\end{equation}
with coefficients $\tilde{\phi}_{nm}({\bf R},t)$ which depend on time $t$ and the atomic positions $r_{i}$, summarized in the vector
${\mathbf R}=(r_1,..,r_N)$. Instead of the states $|\pi_{nm}\rangle$, we may alternatively also use the eigenstates of
the electronic Hamiltonian to span the electronic subspace. These states $|\varphi_k({\bf R})\rangle$ ($k=1,..,2N_AN_B$) generally depend on the atomic positions and satisfy
\begin{equation}
\label{electronic:eigensystem}
H_{\rm el}({\bf R})|\varphi_k({\bf R})\rangle=U_k({\bf R})|\varphi_k({\bf R})\rangle.
\end{equation}
The eigenvalues $U_k({\bf R})$ are often referred to as the adiabatic surfaces. The corresponding expansion of the full wave function can be written as
\begin{equation}\label{fullwfadiab}
\ket{\Psi({\bf R},t)}=\sum_{k}\phi_{k}({\bf R},t)|\varphi_k({\bf R})\rangle.
\end{equation}
We refer to these two possible representations of the wave function as the diabatic (Eq.~(\ref{fullwf})) and adiabatic (Eq.~(\ref{fullwfadiab})) expansion respectively, either of which can be 
a convenient choice, depending on the particular system and/or observable of interest. In our case, the electronic Hamiltonian has the simple structure of. Eq.~(\ref{Hel}) and Table~\ref{tab1},
and as a consequence, the mapping of the diabatic onto the adiabatic basis is quite simple as well. The adiabatic surfaces $U_k({\bf R})$ can be grouped into $N_AN_B$ pairs, each of which can be written in terms of the Hamiltonian 
matrix elements as $U_{nm}^\pm({\bf R})=\pm V_{nm}(r_{nm})$, see Eq.~(\ref{Hel}). The corresponding adiabatic basis states $|\varphi_k\rangle$ become independent of ${\bf R}$ and assume the form
$|\varphi_{nm}^\pm\rangle=(|\pi_{nm}\rangle\pm |\pi_{mn}\rangle)/\sqrt{2}$. The adiabatic expansion of the wave function hence reads
\begin{eqnarray}\label{fullwfadiab_rew}
&&\ket{\Psi({\bf R},t)}=\\
\nonumber
&&\sum_{n\in\mathcal{A},m\in\mathcal{B}}
\left(\phi_{nm}^+({\bf R},t)|\varphi_{nm}^+\rangle + \phi_{nm}^-({\bf R},t)|\varphi_{nm}^-\rangle \right).
\end{eqnarray}
If we insert the above expansion into the time-dependent Schr{\"o}dinger equation, we find that the equations of motion for the coefficients ${\phi}_{nm}^\pm$ decouple for different atom pairs $nm$ and surfaces $\pm$:
\begin{align}
\label{TDSE}
i\hbar \frac{\partial}{\partial t} \phi^{\pm}_{nm}({\bf R},t)&= -\sum_{i=1}^N\frac{\hbar^2}{2M}\nabla^2_{r_i}\phi_{nm}^{\pm} ({\bf R},t)
\CR
&+ U_{nm}^{\pm}(r_{nm}) {\phi}^{\pm}_{nm}({\bf R},t).
\end{align}
Thus, 
%the adiabatic representation in our case has the advantage that 
it is sufficient to consider each atom pair separately in order to obtain the full dynamics.
Moreover, for each atom pair $nm$ the adiabatic surfaces $U_{nm}^{\pm}(r_{nm})$ depend only on the relative interatomic distance. Hence, the time-propagation
can be reduced to a one-dimensional problem by transforming the dynamics for each pair to relative and center of mass coordinates, where the center of mass motion trivially decouples from the dynamics.

Having introduced the representations of the wave functions, we now define the initial state of two superatoms with different angular momenta, 
as mentioned in the introduction. The state at $t=0$ is assumed to be a direct product of the form
\begin{equation}\label{initialPsi}
\ket{\Psi^{\rm ini}}=\ket{\psi_{\rm el}^{\rm ini}}\otimes\ket{\chi_{\rm sp}^{\rm ini}},
\end{equation}
where $\ket{\psi_{\rm el}^{\rm ini}}$ is the initial electronic state and $\ket{\chi_{\rm sp}^{\rm ini}}$ the initial wave function of the atoms.
The former corresponds to all atoms in cloud $A$ coherently sharing the Rydberg $s$-excitation and all atoms
in cloud $B$ coherently sharing the Rydberg $p$-excitation:
\begin{equation}\label{psiel0}
\ket{\psi_{\rm el}^{\rm ini}}=\frac{1}{\sqrt{N_AN_B}}\sum_{\mge{n\in \mathcal{A},\,m\in\mathcal{B}}}|\pi_{nm}\rangle.
\end{equation}
The initial atomic wave function is taken as a product of one-dimensional Gaussians in position space, centered around the respective centers $r_A$, $r_B$ of the two clouds,
\begin{equation}\label{initstate}
\chi_{\rm sp}^{\rm ini}({\bf R})=\prod_{n=1}^{N_A} \chi_{1}^{\sigma}(r_n-r_A)\prod_{n=N_A+1}^{N}\chi_{1}^{\sigma}(r_n-r_B),
\end{equation}
where
\begin{equation}
\label{gaussian}
\chi_{\mathcal{D}}^{\sigma}({\bf x})=\left(\pi\sigma^2\right)^{-\mathcal{D}/4}\exp{(-|{\bf x}|^2/2 \sigma^2 )}
\end{equation}
denotes a Gaussian in $\mathcal{D}$ dimensions.
The width $\sigma$ is taken such that $\sigma<r_{\rm bl}$ and $\sigma\ll L$, where $L=|r_A-r_B|$ is the distance between the clouds.
This choice of the initial spatial wave function models an experimental preparation of the atoms in the lowest oscillator states of two harmonic traps.
From the initial state, we can also directly infer the initial adiabatic expansion coefficients ${\phi}_{nm}^{\pm}(r_{nm},t\! =\! 0)$.
As mentioned before, for each atom pair $nm$ we only consider the relative coordinate $r_{nm}$ in the equation of motion \eref{TDSE} for the adiabatic expansion coefficients,
and the choice \eref{initstate} yields
\begin{equation}\label{initialadwave}
\phi_{nm}^{\pm}(r_{nm},t\! =\! 0)=\frac{1}{\sqrt{2}}\chi_{1}^{\sqrt{2}\sigma}(r_{nm}-L).
\end{equation}
Finally, we note that the position-dependent dipole-dipole potential entangles the electronic interactions and atomic motion~\cite{Moeb_11,Wall_07,Ates_08,Wust_10},
so that for $t>0$ the  direct product form of the initial wave function in~\eref{initialPsi} will not persist.

%%%%%%%%%%%%%%%%%%%%%%%%%%%%%%%%%%%%%%
\section{Dynamics in the electronic subspace}\label{sec3}
%%%%%%%%%%%%%%%%%%%%%%%%%%%%%%%%%%%%%%

\begin{figure}
\begin{footnotesize}
\begin{center}
\scalebox{0.45}{\includegraphics{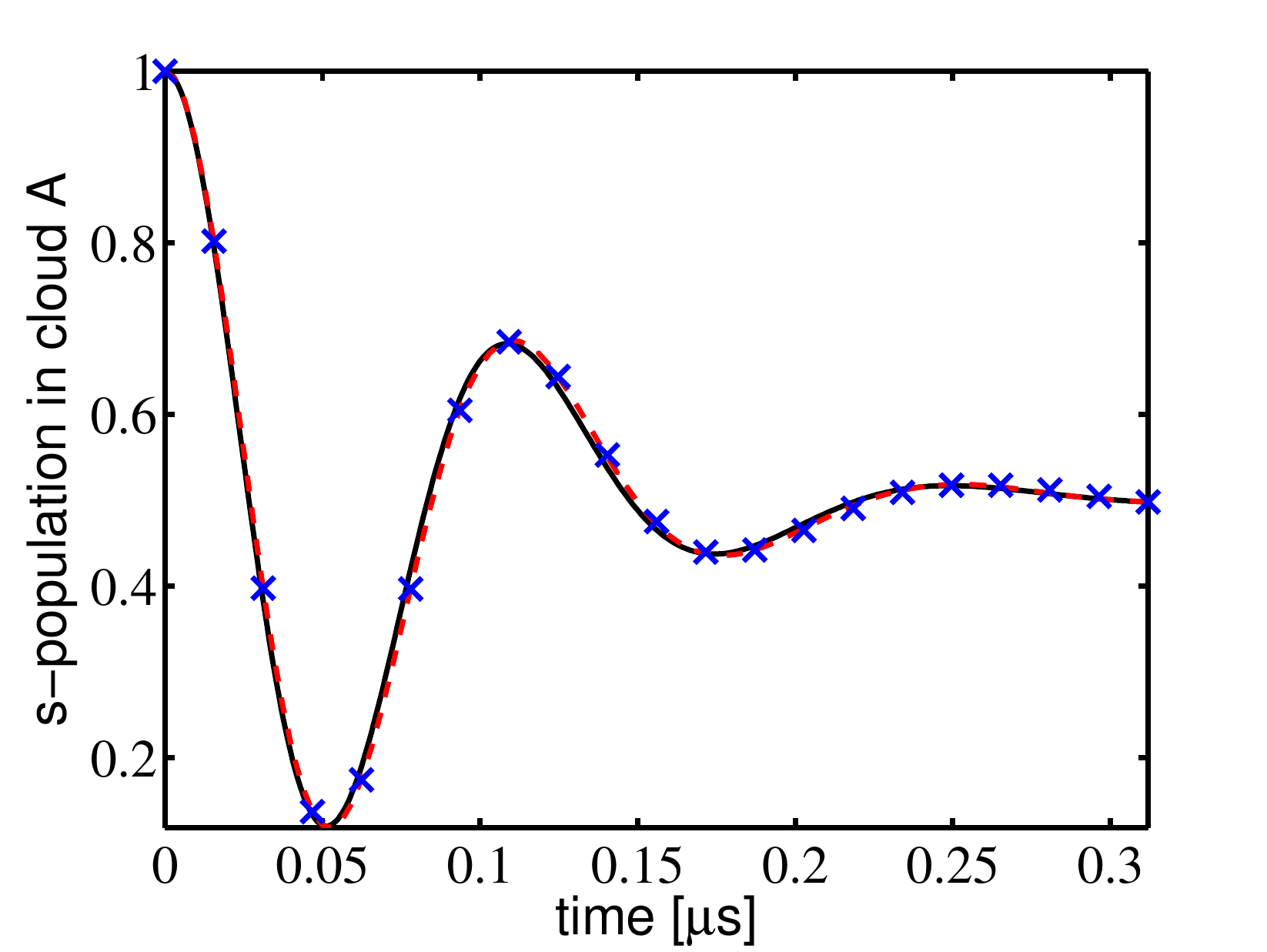}}
\end{center}  
\begin{quote} 
\begin{center}
\caption{
Color online. The solid black line shows the s-population in cloud A for $N_A=N_B=1$ as a function of time, calculated fully quantum mechanically using \eref{TDSE}.
The red dashed line / crosses show the s-population in cloud A calculated as an average over static point particles for $N_A=N_B=1$ / $N_A=N_B=10$, respectively.
The latter is explicitly given in \eref{pbar} and ultimately leads to \eref{analyticdephasing}, see text.
The following parameters were used:
$\sigma=\sqrt{2}/3\,\mu$m, $L=6\,\mu$m, $V_0=10^6\,$a.u., corresponding to Li atoms with the principal quantum number $\nu\approx 30..40$, and
$M=11000\,$a.u. (mass of $^{6}$Li).
}
\label{excitons}
\end{center}
\end{quote}
\end{footnotesize}
\end{figure}

Before tackling the intricate interplay of electronic excitation and atomic motion, we first study the dynamics of the populations of
Rydberg s and p excitations in each cloud caused by the dipole-dipole interactions.
The probability to find an s-excitation on atom $n$ and a p-excitation on atom $m$ is easily obtained from the diabatic representation (cf. Eq.~(\ref{fullwf})):
\begin{equation}\label{elpopu}
P_{nm}(t)=\int d^N {\bf R}\, |\tilde{\phi}_{nm}({\bf R},t)|^2.
\end{equation}
From that, we can define the s-population and p-population in each of the clouds. For example, the s-population in cloud A is given by
\begin{equation}\label{Pquant}
P^s_A(t)=\sum_{n\in \mathcal{A}\,m\in\mathcal{B}}P_{nm}(t).
\end{equation}
It is sufficient to consider $P^s_A(t)$, since the s-population in cloud B and the p-population in each cloud, defined in the same manner, directly follow
from the conservation of the total probability.
For our initial state (Eq.~(\ref{psiel0})) we have $P^s_A(t\! =\! 0)=1$. Due to the interactions (Eq.~(\ref{dipdip})), 
the s-population will migrate from cloud $A$ to cloud $B$, while p-population will migrate from cloud $B$ to cloud $A$, resulting in
Rabi-oscillations of the quantity $P^s_A(t)$. These oscillations dephase due to the width of the initial
spatial wave functions $\ket{\chi_{\rm sp}^{\rm ini}}$ and can be quite accurately described by the following analytic expression:
\begin{equation}\label{analyticdephasing}
P^{s}_{A}(t)\approx\frac{1}{2}\left[1+\cos(\omega t)\exp\left(-\left(\frac{3\sigma}{\sqrt{2}L}\omega t\right)^2\right)\right],
\end{equation}
with $\omega=(2V_0/L^3)/\hbar$.
As shown in Fig.~\ref{excitons}, this expression indeed shows very good agreement with the exact quantum mechanical result.

To illustrate the origin of this behavior, let us start with a very simple picture and treat the atoms as point particles fixed in space.
In that case, the entire dynamics is encapsulated by the wave function in the electronic subspace,
\begin{equation}
\label{fullwfdisc}
\ket{\Psi_{\rm el}(t)}=\sum_{nm}f_{nm}(t)|\pi_{nm}\rangle,
\end{equation}
with the coefficients $f_{nm}$ evolving according to
\begin{align}   
\label{TDSEdiscrete}
i\hbar \frac{\partial}{\partial t} f_{nm}(t)&=V_{nm}(r_{nm}) f_{mn}(t).
\end{align}
Consequently, the s-population in cloud A in the fixed point particle limit $\mathcal{P}^s_{A}(t)$ is given by
\begin{equation}\label{Pclass}
\mathcal{P}^s_{A}(t)=\sum_{n\in \mathcal{A}\,m\in \mathcal{B}}|f_{nm}(t)|^2.
\end{equation} 
To obtain an explicit expression for this quantity, we first consider the simplest case $N_A=1$ and $N_B=1$, where we denote the distance of the two atoms by $d$.
The probability $p(t)$ to find the s--excitation at time $t$ on the atom in cloud $A$ is easily obtained from \eref{TDSEdiscrete}:
\begin{equation}
p(t)=|\langle \pi_{12} | \Psi_{\rm el}(t) \rangle|^2=|f_{12}(t)|^2=\cos^2\left(\frac{V_0 t}{d^3\hbar}\right).
\end{equation}
The expression in \eref{Pclass} is nothing but the average of $p(t)$ over all the atom pairs $nm$. In other words, we consider a large number of realizations $k$ of single atom pairs
with interatomic distances $d_k$ and define
\begin{equation}
p_k(t)=\cos^2\left(\frac{V_0 t}{d_k^3\hbar}\right).
\end{equation}
This yields 
\begin{equation}\label{pbar}
\mathcal{P}^s_{A}(t)=\frac{1}{G}\sum_{k=1}^{G}p_{k}(t)
\end{equation}
for the s-population in cloud A in the fixed point particle limit, where $G$ is the total number of atom pairs.
The expression does not change for the general case $N_A>1$, $N_B>1$: From the block structure of the electronic Hamiltonian, one can infer that the averaged result is exactly the same as in the case
$N_A=1$ and $N_B=1$, we simply have added to the average over positions an average over blocks, which has exactly the same structure.
%%Using a trigonometric theorem we can rewrite the expression as
%%\begin{equation}
%%\mathcal{P}^s_{A}(t)=\frac{1}{2}+\frac{1}{2G}\sum_{k=1}^{G}\cos(2V_{0}t/d_{k}^{3}\hbar).
%%\end{equation}
In the limit $G\rightarrow\infty$, the sum in \eref{pbar} becomes an integral, in which the distances $d_k$ follow the spatial distribution of the fixed atoms.
If the classical spatial distribution of the distances, $F(d)$, is chosen as $F(d)=|\chi_{1}^{\sqrt{2}\sigma}(d-L)|^2$, we arrive at the integral
\begin{equation}\label{theintegral}
\mathcal{P}^s_{A}(t)=\left(\frac{2}{\pi\sigma^2}\right)^{1/2}\int {\rm d}L'\,\cos^2\left(\frac{V_0 t}{\hbar L'^3}\right)\exp\left[-\frac{(L'-L)^2}{2\sigma^2}\right].
\end{equation}
This integral cannot be solved analytically exactly, but by substituting $\eta=L-L'$ and Taylor-expanding $(L+\eta)^{-3}\approx L^{-3}(1-3\eta/L)$
one finds \eref{analyticdephasing}. 

It may seem surprising that a simple model which ignores both the atomic wave functions and the interplay between electronic excitation and atomic motion
gives a correct result. This is due to the fact that the time scales on which the exchange of excitation population occurs are very short compared to the time when atomic motion becomes relevant, and
the kinetic energy on these timescales is very small compared to the dipole-dipole interaction. 
In \aref{pointparticle:appendix} we explain this agreement more formally using a semiclassical propagator and demonstrate in particular why an average over a static ensemble is able to resemble
the quantum mechanical result.

To summarize this section, we conclude that due to the particular form of $H_{\rm el}$, the time evolution of electronic populations per cloud for arbitrary $N_A$, $N_B$ is identical to the average over an ensemble of systems 
with $N_A=N_B=1$. 
We explicitly demonstrated this using a point particle model, and further argue in \aref{blockdiagonality} that this behaviour will persist fully quantum mechanically.
Note, however, that while the point particle approximation gives a correct description for the specific case studied here, it will in general give results different from a full quantum mechanical treatment.
As an example, consider a setup in which the two clouds are additionally confined in two strong harmonic traps with energy spacing $\Delta E\gg V_0/L^3$.
In this case the kinetic energy term is by construction much larger than the dipole-dipole interaction.
Then, in a quantum treatment, $P^s_A(t)$ would undergo long coherent oscillations without dephasing, while an average over point particles would show fast dephasing as without the traps.

%%%%%%%%%%%%%%%%%%%%%%%%%%%%%%%%%%%%%%
\section{Atomic motion}\label{sec4}
%%%%%%%%%%%%%%%%%%%%%%%%%%%%%%%%%%%%%%

We now consider timescales on which the dipole-dipole-induced atomic motion becomes relevant. 
%As will be seen in this section, the fact that
%the time evolution of electronic populations for any atom number is equivalent to an average over single pairs gives rise to quite
%interesting motional dynamics.
Although the motion of the atoms is restricted to one dimension along the 
separation of the clouds, the full many-body problem is still too complex to be treated fully quantum mechanically.
Therefore, we use a quantum-classical 
hybrid approach, which has been successfully applied to similar systems~\cite{Ates_08,Wust_10,Moeb_11}, namely Tully's fewest switching 
algorithm~\cite{Tull_71,Tull_90,Hamm_94}. The method relies on a classical treatment of the atomic trajectories 
and a quantum mechanical treatment of the electronic degrees of freedom.
A detailed description can be found in the aforementioned references, and we only briefly summarize 
the main aspects relevant for the present problem in Appendix~\ref{Tully}. 

\begin{figure}
\begin{footnotesize}
\begin{center}
\scalebox{0.38}{\includegraphics{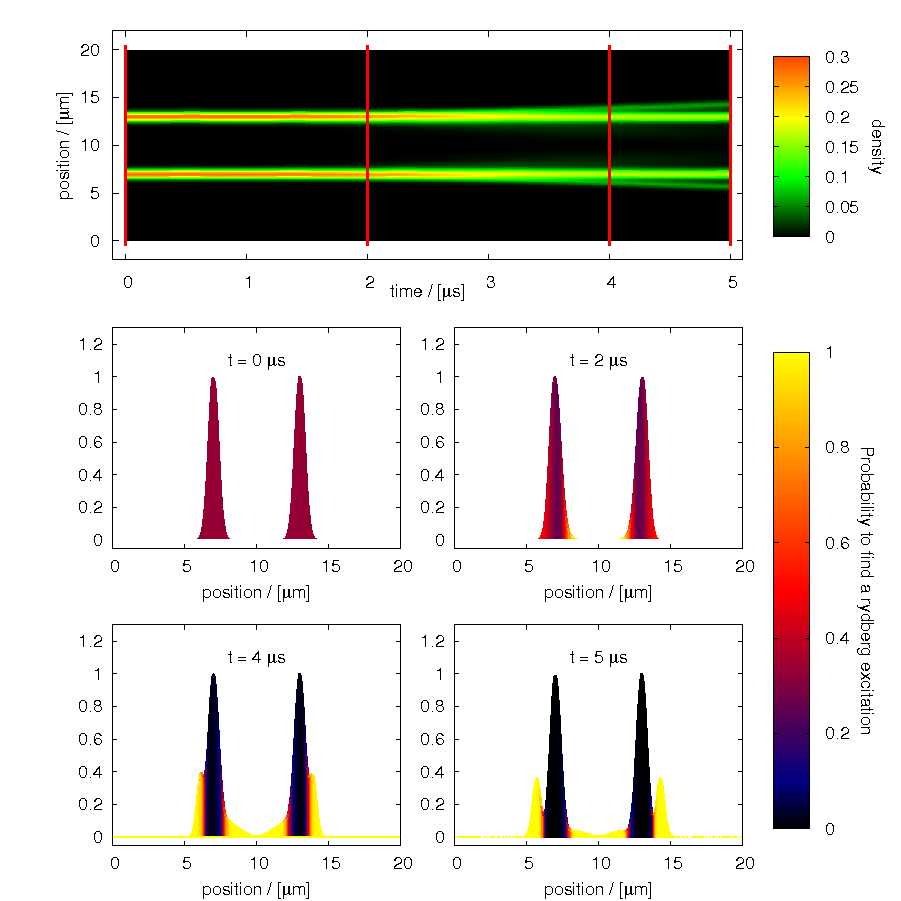}}
\end{center}
\begin{quote}
\caption{
Color online. Atomic motion induced by dipole-dipole interactions. The upper panel shows the atomic density as a function of time.
The lower four panels show snapshots of the density at times $t=0,2,4,5\,\mu$s. The color indicates the probability
to find any Rydberg excitation ($|s\rangle$ or $|p\rangle$) at a given position. In this figure, we do not distinguish
between $|s\rangle$ and $|p\rangle$ excitations, since the dephasing oscillations of electronic population,
as shown in Fig.~\ref{excitons}, occur on the short timescale $t\sim 0.2\,\mu$s and are therefore not shown here.
We observe that the main bulk of the atoms remains at rest,
with vanishing probability to find an excitation. A pair of single atoms is ejected, and the Rydberg excitation is entirely 
localized on those two atoms. The calculations are shown for $N_A=N_B=3$, for the same parameters as in Fig.~\ref{excitons}
but with $M=12600\,$a.u., corresponding to the mass of $^{7}$Li. 
}
\label{figmotion}
\end{quote}
\end{footnotesize}
\end{figure}

The results of the simulations are shown in Fig.~\ref{figmotion}. We observe that, although the initial Rydberg excitation
is coherently shared by all atoms within one cloud, it is nevertheless just a single atom that is ejected from each cloud.
\sew{The atoms are still in a coherent superposition where \emph{each of them} has left its cloud with some probability, but the ejected atoms always carry the excitation and an ensemble of ground-state atoms stays behind.  
In this sense the superatoms break apart when subject to dipole-dipole induced motion, since excited and ground-state atoms are now spatially} \mge{separated.}

Another interesting aspect is that the resulting motion is a superposition of repulsion and attraction.
It is known~\cite{Ates_08} that for single Rydberg atoms interacting via dipole-dipole-forces, the direction of the
force can be controlled via the initial electronic state; The same holds for our system, i.e., the choice of 
$\ket{\psi_{\rm el}^{\rm ini}}$ in Eq.~(\ref{psiel0}) determines the atomic motion. For $V_0>0$, the initial electronic states
\begin{equation}\label{psirepatt}
\ket{\psi_{\rm el}^{\pm}}=\frac{1}{\sqrt{2N_AN_B}}\sum_{\mge{n\in \mathcal{A},\,m\in \mathcal{B}}}\left(|\pi_{nm}\rangle\pm|\pi_{mn}\rangle\right)
\end{equation}
correspond to purely attractive (+) and repulsive (-) motion of the atom pair. The initial state chosen in Eq.~(\ref{psiel0}),
describing two superatoms, is a superposition of these two states, and consequently leads to a superposition of two opposite
motion directions.

\begin{figure}
\begin{footnotesize}
\begin{center}
\scalebox{0.2}{\includegraphics{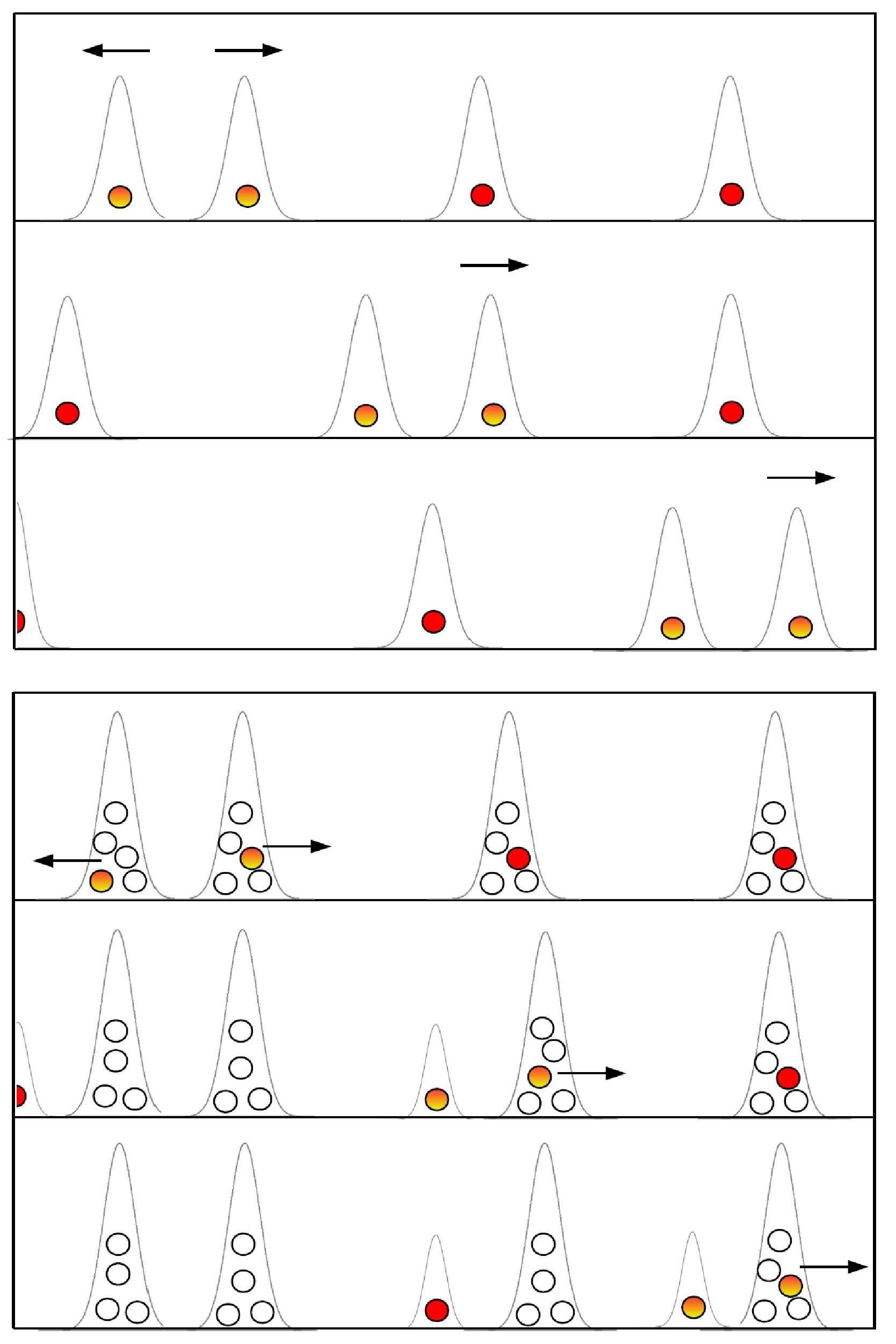}}
\end{center} 
\begin{quote}
\caption{Color online. Sketch of the excitation transport scheme with atom clouds as sites (lower panel), compared to the one with single atom sites (upper panel) as demonstrated in Refs.~\cite{Wust_10, Moeb_11}.
Color codes are as in Fig.~\ref{fig0}.
All but one atom from each cloud remain at rest and do not participate in the dynamics, hence excitation and entanglement transport occur equivalently in both schemes.
For the sake of clarity, the lower panel shows the dynamics \sew{for a well defined choice of initially excited atoms in each cloud, while the complete dynamics consists of a superposition where each atom in each cloud has the chance to be the only one from that cloud participating in the dynamics.}
\label{cloudtransport}}
\end{quote}
\end{footnotesize}
\end{figure}

\begin{figure}
\begin{footnotesize}
\begin{center}
\scalebox{0.22}{\includegraphics{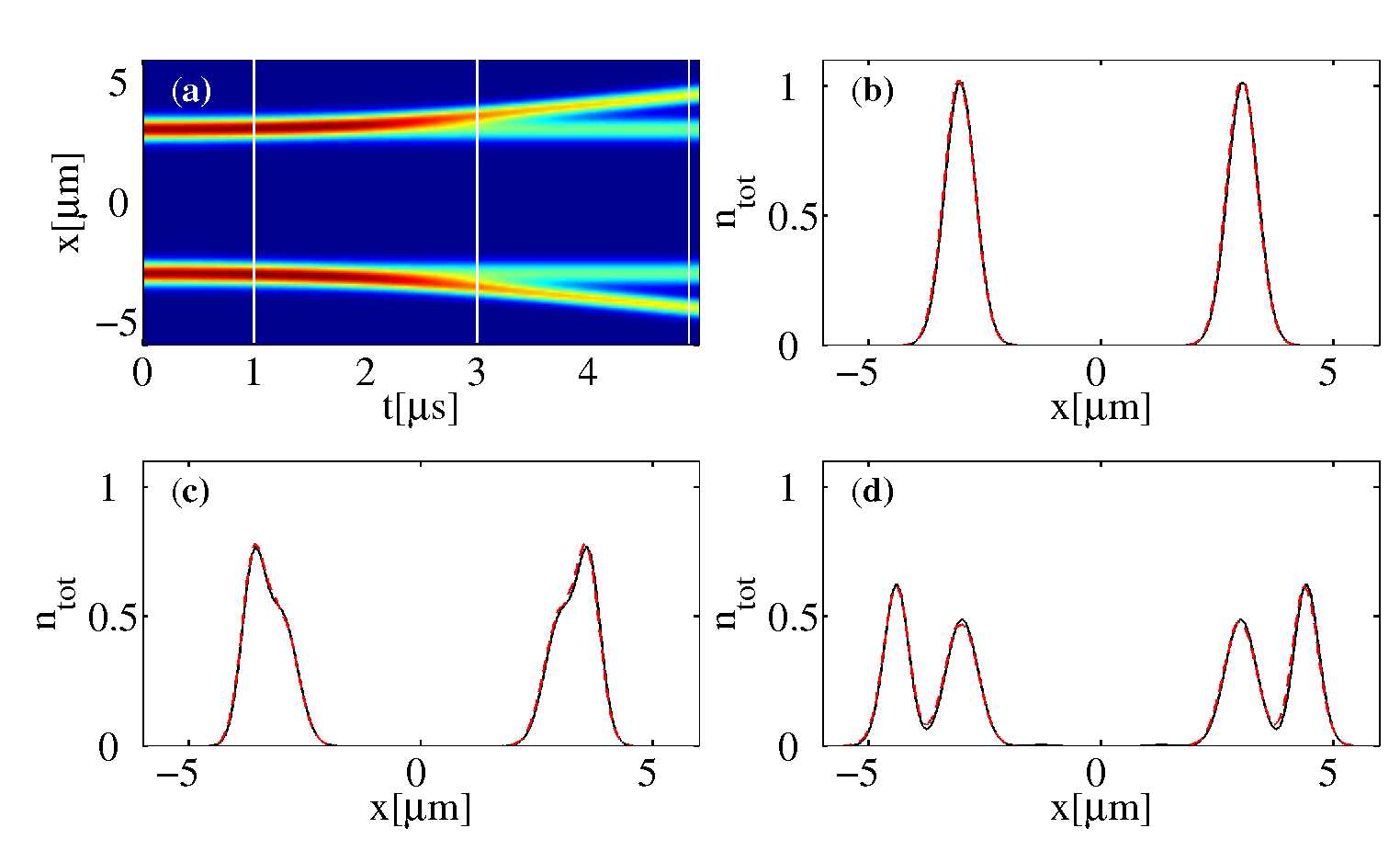}}
\end{center}
\begin{quote}
\caption{
Color online. Comparison of the atom density \sew{during superatom break-up}, obtained from a full quantum mechanical calculation \sew{using \eref{TDSE}} (solid black line) and Tully's 
quantum-classical hybrid algorithm (red dashed line). (a) Total atom density as a function of time. (b)-(d): 
Snapshots at three different times, as indicated by white vertical lines in (a): $t_1=1\,\mu$s (b), $t_2=3\,\mu$s (c), $t_3=4.9\,\mu$s (d). The same parameters as in Fig.~\ref{excitons}
were used. For the sake of clarity, the comparison is shown for an initially purely repulsive electronic state, cf. Eq.~(\ref{psirepatt}).
}
\label{quantumTully}
\end{quote}
\end{footnotesize}
\end{figure}

The presented results have interesting implications for quantum transport protocols based on single-atomic aggregates, 
such as adiabatic entanglement transport~\cite{Wust_10,Moeb_11}. Our findings show that even in an ensemble of atoms
which share the Rydberg excitation, only one of them will be set in motion by dipole-dipole forces. This facilitates
the experimental realization of transport schemes in~\cite{Wust_10,Moeb_11}, since there is apparently no need to isolate single
atoms. As formally discussed in Appendix~\ref{blockdiagonality}, our results can be extended to more than two clouds, due to the block structure of the full Hamiltonian, Eq.~(\ref{fullHam}),
and the corresponding time evolution operator $U(t)=\exp{[-i Ht]}$.
A 'Gedankenexperiment' presenting the transport scheme of Refs.~\cite{Wust_10,Moeb_11} both,
with single atoms and with atoms clouds, is sketched in Fig.~\ref{cloudtransport}. 
If atom clouds are used one should consider that the transport efficiency may be affected by collisions of moving Rydberg atoms
with ground state atoms at rest for a large number of atoms per cloud. However, as estimated in Appendix~\ref{motion_background}, in an ultracold gas such processes can be neglected on the microsecond timescale.
Let us finally mention that the validity of the adopted quantum-classical hybrid approach was verified by comparing the time
evolution of the spatial probability density with a full quantum calculation for the accessible case $N_A=N_B=2$, as demonstrated in Fig.~\ref{quantumTully}.
%

%%%%%%%%%%%%%%%%%%%%%%%%%%%%%%%%%%%%%%
\section{Summary and conclusions}\label{sec6}
%%%%%%%%%%%%%%%%%%%%%%%%%%%%%%%%%%%%%%

We have studied the effects of dipole-dipole interactions on two Rydberg-blockaded atom clouds. Each cloud is initially prepared in a superatomic state, where the atoms within one cloud coherently share
a Rydberg s- and p-excitation, respectively.
On short timescales, we observe dephasing oscillations in the angular momentum of the Rydberg excitations in both clouds. We found that the dephasing many-body system can be resembled by an average over single
dipole-dipole-interacting atom pairs, as long as the spatial distribution of these pairs mimics the initial quantum mechanical density.
On long timescales, the dipole-dipole interactions induce atomic motion. For the setup considered in this paper, the forces physically remove
the initial coherent single excitation from both clouds. A single atom pair leaves the clouds with the excitation 
entirely localized on this pair. This finding may facilitate an experimental realization of quantum transport protocols in Rydberg aggregates, 
such as proposed in~\cite{Mulk_07,Wust_10} since one-atom-sites can be replaced by micro traps containing several atoms.

%%%%%%%%%%%%%%%%%%%%%%%%%%%%%%%%%%%%%%
\appendix
%%%%%%%%%%%%%%%%%%%%%%%%%%%%%%%%%%%%%%

%%%%%%%%%%%%%%%%%%%%%%%%%%%%%%%%%%%%%%
\section{Semiclassical propagation on short time scales}
\label{pointparticle:appendix}
%%%%%%%%%%%%%%%%%%%%%%%%%%%%%%%%%%%%%%

In this section, we re-derive the expression from \eref{analyticdephasing} using the semiclassical Herman-Kluk-Propagator~\cite{kluk:semiclassexample,grossmann:inivalueprop,Gross_02,Gross_femtosecond}.
We have already seen in section~\ref{sec3} that the expression correctly describes the dephasing Rabi-oscillations of electronic populations, since it agrees with a full quantum mechanical 
calculation (see Fig.~\ref{excitons}). We have also seen that it can be obtained within a rather crude approximation, namely by modeling the atoms as fixed point particles.
The following derivation serves as an explanation why this approximation works in our case.

The Herman-Kluk propagator $K^{HK}$ evolves a quantum mechanical wave function in time semiclassically by means of classical phase-space trajectories. Given an initial wave function $\phi(\bv{R}',t\! =\! 0)$
at time $t=0$ and position ${\bf R}'$, the propagator provides the wave function at time $t$ and position ${\bf R}''$:
\begin{equation}\label{whatHKdoes}
\phi(\bv{R}'',t)=\int d{\bf R'} K^{HK}(\bv{R}'',t;\bv{R}',t\! =\! 0){\phi}(\bv{R}',t\! =\! 0).
\end{equation}
We are going to apply the propagator to the adiabatic wave functions $\phi^{\pm}_{nm}({\bf R},t)$ (cf. Eqs.~(\ref{TDSE}) and~(\ref{initialadwave})). 
We will omit the subscript $nm$ in this section, and further restrict the treatment to a single coordinate 
$r_{nm}\equiv r$. As already argued in section~\ref{sec2}, this is sufficient due to the decoupling of the equations of motion for each atom pair and the transformation of the problem onto relative and center of mass 
coordinates.

The explicit form of the propagator is given by
\begin{align}
\label{semiclassprop}
&K^{HK}(R'',t;R',t\! =\! 0)^{\pm}=
\CR
& \int \frac{dpdq}{(2\pi\hbar)}\braket{R''}{z_t}  C({q}_t,{p}_t,{q},{p})
\CR
 &\times \exp{\left( \frac{i}{\hbar}S^{\pm}[{q}_t,{p}_t] - \frac{i}{2\hbar}  ({q}_t{p}_t -{q}{p})  \right)}\braket{z}{{R}'}.
\end{align}
Here, $\ket{z}$ are coherent states centered on position ${q}$ and momentum ${p}$,
\begin{align}
\label{cohst}
\braket{{R}}{z} &= \chi^{\sigma_0}_{1}({R} - {q})\exp\left[\frac{i}{\hbar} {p} \left({R} - \frac{{q}}{2}\right)\right],
\end{align}
with $\chi^{\sigma}_{N}$ from Eq.~(\ref{gaussian}). The width $\sigma_0$ of these states is just a parameter that is relevant in a numerical implementation and hence shall not concern us here.
The classical trajectories (${q}_t$, ${p}_t$) 
evolved with the Hamiltonian $\sub{H}{c} = {p}_t^2/2M + U^{\pm}({q}_t)$ from the initial conditions (${q'}$, ${p'}$) accumulate the action 
$S^{\pm}[{q}_t,{p}_t]=\int [{p}_t^2/2M - U^{\pm}({q}_t)]dt$. The prefactor $C$ is given by 
\begin{align}
\label{prefactor}
C&= \frac{1}{2}\left( \pdiff{{p}_t}{{p}} +  \pdiff{{q}_t}{{q}}  - \frac{i\hbar}{\sigma_0^2}   \pdiff{{q}_t}{{p}}  -  \frac{\sigma_0^2}{ i\hbar }\pdiff{{p}_t}{{q}} \right)^{\frac{1}{2}},
\end{align}
which is a complex number. Note that in the multidimensional case, the prefactor is a complex valued determinant composed of stability matrix blocks~\cite{Gross_femtosecond}.

We consider short times and hence expand the classical trajectories (${q}_t$, ${p}_t$) up to first order in time,
\begin{eqnarray}\label{analtrajs}
{q}_t&=&{q}+\frac{p}{m}t,
\\
{p}_t&=&{p} - \frac{\partial U^{\pm}({q})}{\partial {q}}t.
\end{eqnarray}
%and neglecting all higher orders in $t$ here and in the following. 
This is, in fact, the central approximation which enters the present derivation.
It simplifies \eref{prefactor} to
\begin{equation}\label{prefac2}
C=C(q)=\frac{1}{2}\left(2-\frac{i\hbar}{\sigma_0^2}\frac{t}{M}+\frac{\sigma_0^2}{i\hbar}\frac{\partial^2 U^{\pm}({q})}{\partial{q}^2}t\right)^{1/2}.
\end{equation}
After inserting the trajectories into the expression for the Herman-Kluk propagator, expanding $U^{\pm}({q}_{t'})$ around $t'=0$ to first order in time and collecting all terms,
we obtain
\begin{eqnarray}
&&K^{HK}(R'',t;R',t\! =\! 0)^{\pm}= \\
\nonumber
&&\int \frac{dpdq}{(2\pi\hbar)}C(q)\chi^{\sigma_0}_{1}({R'} - {q})\chi^{\sigma_0}_{1}({R''} - {q})\\ 
\nonumber
&&\times \exp\left[\frac{i}{\hbar} {p} \left({R''} - R'\right)\right]\\
\nonumber
&&\times \exp{\left[-\frac{it}{\hbar}\pdiff{{U^{\pm}(q)}}{{q}}(R''-{q})\right]}\exp{\left[-\frac{it}{\hbar}U^{\pm}(q)\right]}.
\end{eqnarray}
Carrying out the momentum integration yields a $\delta$-function in $R'$ and $R''$, 
\begin{eqnarray}\nonumber
&&K^{HK}(R'',t;R',t\! =\! 0)^{\pm}= \int dq\,C(q)\left[\chi^{\sigma_0}_{1}({R'}-{q})\right]^2 \\
\nonumber
&&\times\exp\left[-\frac{it}{\hbar}\left(U^{\pm}({q})+
\frac{\partial U^{\pm}({q})}{\partial{q}}({R}'-{q})\right)\right]\delta({R'}-{R''}).\\
\end{eqnarray}
The delta-function in the propagator tells us that we can indeed map the quantum mechanical result onto an average over static realizations.
Finally, to find the explicit form of this mapping, we have to
perform the integration in position space. In the semiclassical context of $K^{HK}$ this is done consistently with the
stationary phase approximation~\cite{Gross_femtosecond}:
\begin{eqnarray}
&&\int_{-\infty}^{\infty}dx\,\exp(if(x)/\eta)g(x)\\
\nonumber
&&\underset{\eta\rightarrow 0}{=}\sqrt{\frac{2\pi i\eta}{f''(x_0)}}\exp(if(x_0)/\eta)g(x_0),
\end{eqnarray}
where $x_0$ is the stationary point satisfying $f'(x_0)=0$.
Using this formula and the $\hbar\rightarrow 0$ limit of the prefactor $C(q)$,
we arrive at the following simple expression for the Herman-Kluk propagator:
\begin{equation}
K^{HK}(R'',t;R',t\! =\! 0)^{\pm}=\exp\left[-\frac{it}{\hbar}U^{\pm}({R'})\right]\delta({R'}-{R''}).
\end{equation}
By applying this propagator to the initial adiabatic expansion coefficients from \eref{initialadwave} according to \eref{whatHKdoes}
we obtain the semiclassical time evolution in the adiabatic picture. From that, we can immediately extract the semiclassical limit
for, say, the s-population in cloud A as
\begin{equation}
P^{sc}_{s,A}(t)=\int {\rm d}r_{nm}\left|\tilde{\phi}_{nm}(r_{nm},t)^{+}+\tilde{\phi}_{nm}(r_{nm},t)^{-}\right|^2.
\end{equation}
After insertion of the Herman-Kluk-propagated wave functions, the expression reads
\begin{equation}
P^{sc}_{s,A}(t)=\left(\frac{2}{\pi\sigma^2}\right)^{1/2}\int {\rm d}L'\,\cos^2\left(\frac{V_0 t}{\hbar L'^3}\right)\exp\left[-\frac{(L'-L)^2}{2\sigma^2}\right],
\end{equation}
which is identical to \eref{theintegral}. 

%%%%%%%%%%%%%%%%%%%%%%%%%%%%%%%%%%%%%%
\section{Atomic motion from block-diagonal electronic Hamiltonians}
\label{blockdiagonality}
%%%%%%%%%%%%%%%%%%%%%%%%%%%%%%%%%%%%%%
In Section~\ref{sec3}, we have seen that
the electronic Hamiltonian for a collection of Rydberg-blockaded atom clouds, Eq.~(\ref{Hel}), can be cast into a block-structure, where in each block only a single atom per cloud participates in non-trivial electronic dynamics. In 
this appendix we formally show that this leads to 
rapid entanglement of 
motion and electronic state, since in each block also only one atom per cloud performs non-trivial motional dynamics. 
This holds for any number of clouds and atoms per cloud.
We show that non-adiabatic transitions do not affect this picture, as they respect the block structure. 
Let us consider the example of three clouds with two atoms each. The generalization to other numbers is straightforward.

Let the atoms be grouped into clouds $\mathcal{A}=\{1,2\}$,  $\mathcal{B}=\{3,4\}$,  $\mathcal{C}=\{5,6\}$.
In that case the total Hamiltonian, including atomic motion and the electronic degrees of freedom, can be written as
\begin{equation}\label{Hblocks}
H=\sum_{i=1}^6{T}_n + \mbox{diag}[H_1,H_2,H_3,...].
\end{equation}
We have abbreviated the kinetic energy term for each atom as $T_n = -\hbar^2\nabla^2_{r_n}/(2M)$. 
The symbol $\mbox{diag}[H_1,H_2,H_3,...]$ denotes a block-diagonal matrix, with $3\times3$ blocks $H_j(r_k,r_l,r_m)$. Crucially, each block only depends on the coordinates of \emph{three} atoms, 
with $k\in \mathcal{A}$, $l\in \mathcal{B}$, $m\in \mathcal{C}$, i.e., one from each cloud. The latin index $j$ numbers the blocks. 
Consequently, the time-evolution operator $\hat{U}(0,t)=\exp{[-i Ht/\hbar]}$ also assumes a block-structure $\hat{U}=\mbox{diag}[\hat{U}_1,\hat{U}_2,\hat{U}_3,...]$, with 
\begin{align}
\label{Ublockstructure}
\hat{U}_j&= \exp{[-i (\sum_{n\neq\{k,l,m\}  }T_n)t/\hbar]}
\CR
&\times \exp{[-i(T_k + T_l + T_m + H_j(r_k,r_l,r_m))t/\hbar]}.
\end{align}
\sew{We have used the operator hat on $\hat{U}$ here exclusively to avoid confusion with the potential energy surfaces $U$ occuring elsewhere.}

In \eref{Ublockstructure}, the second exponential acts only on atoms $k$,  $l$,  $m$, while the first one describes free motion of the remaining atoms. 
Now, consider a time evolution beginning in an eigenstate of the electronic Hamiltonian $H_{\rm el}$ which will have support only in a single block $j_0$ of $H$. The 
form of $\hat{U}_j$ immediately tells us that three particles undergo dipole-dipole dynamics and the other ones perform free motion.  
Starting in a non-eigenstate will lead to quick entanglement of motion and electronic state, as in each block three particles with \emph{different} $j$ participate in non-trivial dynamics. The block structure of 
$\hat{U}_j$ also immediately confirms the absence of non-adiabatic transitions between exciton states in different 
blocks. For an alternative argument, note that $\nabla H_{\rm el}$ will have the same block-diagonal structure as $H_{el}$. From that we can see that non-adiabatic coupling terms (defined in the next 
appendix~\ref{Tully} in Eq.~\ref{nonadcoup}) vanish whenever the involved adiabatic eigenstates have support in different blocks.
%\textcolor{red}{Hier ggf noch 2:2:2 Bild}.

%%%%%%%%%%%%%%%%%%%%%%%%%%%%%%%%%%%%%%
\section{Tully's algorithm}
\label{Tully}
%%%%%%%%%%%%%%%%%%%%%%%%%%%%%%%%%%%%%%
We start from the total Hamiltonian as given in Eqs.~(\ref{fullHam}) or~(\ref{Hblocks}). Tully's algorithm is implemented as follows.
One first finds the adiabatic eigenstates in the electronic subspace $|\varphi_k({\bf R})\rangle$ and the corresponding energies $U_k({\bf R})$, and expands the
wave function in this basis, cf. Eqs.~(\ref{electronic:eigensystem}) and~(\ref{fullwfadiab}). However, contrary to a full quantum mechanical approach, the expansion
coefficients do not depend on the atomic positions, and here we denote them by $c_k(t)$ to emphasize the difference,
\begin{equation}
\Psi({\bf R},t)=\sum_{k=1}^{2N_AN_B}c_k(t)|\varphi_k({\bf R})\rangle.
\end{equation}
The adiabatic expansion is inserted into the time-dependent Schr{\"odinger} equation, which leads to a set of equations for the time-dependent expansion
coefficients,   
\begin{equation}\label{TullyClass}
i\hbar \dot{c}_k=U_kc_k-i\hbar \sum_{j=1}^{2N_AN_B}\dot{\bf R}\cdot {\bf d}_{kj}c_j.
\end{equation}
In \eref{TullyClass}, we introduced non-adiabatic coupling vectors
\begin{equation}\label{nonadcoup}
{\bf d}_{kj}=\langle\varphi_k|\nabla_{\bf R}\varphi_j\rangle=\frac{\langle\varphi_k|\nabla_{\bf R}H_{\rm el}({\bf R})|\varphi_j\rangle}{U_j({\bf R})-U_k({\bf R})}.
\end{equation}
Atomic motion is treated classically, i.e., the time evolution of the spatial degrees of freedom is obtained from Newton's equations of motion,
\begin{equation}
M\ddot{\bf R}=-\nabla_{\bf R}U_k({\bf R}),
\end{equation}
and averaged over many trajectories.
The averaging is performed such that the initial conditions for Newton's equations, namely the position ${\bf R}(t=0)$, the velocity ${\bf \dot{R}}(t=0)$ and the adiabatic surface $k(t=0)$,
resemble the Wigner distribution of the initial state.
Then each trajectory is propagated
on a single adiabatic surface $k$ and the presence of non-adiabatic couplings ${\bf d}_{kj}$ is accounted for by introducing the possibility for instantaneous stochastic
switches to another adiabatic surface $j$, as described in detail in Ref.~\cite{Tull_90}.

%%%%%%%%%%%%%%%%%%%%%%%%%%%%%%%%%%%%%%
\section{Rydberg atoms moving through background gas}\label{motion_background}
%%%%%%%%%%%%%%%%%%%%%%%%%%%%%%%%%%%%%%
In Section~\ref{sec4}, we argue that excitation and momentum transport in Rydberg aggregates, as described, e.g., in Refs.~\cite{Wust_10,Moeb_11}, should, in principle, be equally possible 
if single atom sites are replaced by atom clouds. To corroborate this claim, in what follows we estimate the impact of the processes that could potentially spoil a dynamics such as sketched in Fig.~\ref{cloudtransport}.
Let us consider Alkali atoms in an ultra-cold gas. They can be routinely excited to Rydberg states, and in general collisions with surrounding atoms do 
not necessarily lead to a loss of the excited state~\cite{Kano_85,Walz_04,Choi_05}.
However, in the setup considered in Section~\ref{sec4}, some Rydberg atoms are additionally accelerated within the gas due to the dipole-dipole interactions, with typical velocities of the order of $v_m=4\,$m/s for the example of 
$^7$Li. This is about two orders of magnitude larger than the thermal atomic velocity in an ultra-cold $^7$Li gas at $T=1{\mu}K$ of $v_T=4.8\,$cm/s. 
Yet, as we estimate below, this does not lead to new regimes of atomic collisions.
We consider the impact of the following processes:  
(i) Inelastic collisions of Rydberg and ground state atoms, (ii) elastic collisions of Rydberg and ground state atoms, and (iii) collisional ionization of Rydberg atoms.

(i) We can estimate the rate $\Gamma_q$ of Rydberg state quenching collisions with background atoms, 
using $\Gamma_q = v \rho \sigma_q(\nu)$, where $v$ is the velocity of the Rydberg atom with principal quantum number $\nu$, $\rho$ is the density of background atoms and $\sigma_q(\nu)$ is the cross section for 
the process. From \cite{Hugo_83} we have, e.g., for Rubidium atoms, $\sigma_q(40)=2\times 10^{-11}\,{\rm cm}^2$. At $v=4\,{\rm m}/s$ 
and with density $\rho=1\times 10^{18}\,{\rm m}^{-3}$ the inverse quenching rate is $\tau =1/\Gamma_q=125\,\mu$s, which is still larger than the Rydberg state lifetime 
$\tau=46\mu$s. Note that for a scheme as in Fig.~\ref{cloudtransport} the effective background density would be an average over the intra- and intercloud regions and
thus much smaller than the cloud peak density.

(ii) For elastic collisions, Ref.~\cite{Kaul_87} gives a cross section of the order of $\sigma_{\rm el}\approx 10^4\,{\rm a.u.}\approx0.25\times 10^{-12}\,{\rm cm}^2$, 
which is about a factor of 80 smaller than the one given above for inelastic collisions.
Elastic collisions are therefore even less important for the considered range of parameters.

(iii) In order to estimate the order of magnitude for the collisional ionisation rate, we adopt an approximative formula derived by Lebedev~\cite{Lebe_91}. For $\nu\gg l$, it reads, in atomic units,
\begin{equation}
\Gamma_{\rm ion}=\frac{8\sigma_{\rm el}T\nu_{\rm eff}}{\pi\mu}\exp\left(-\frac{1}{2\nu^2_{\rm eff}T} \right),
\end{equation}
where $\sigma_{\rm el}$ is the cross section for elastic scattering of a Rydberg electron on a ground state atom, $\mu$ the reduced mass of the colliding system, 
$T$ the temperature and $\nu_{\rm eff}=\nu-\delta_l$, $\delta_l$ being the quantum defect.
In the ultracold regime, the low temperature of $\sim 1\mu{\rm K}\approx 0.3\times 10^{-11}\,{\rm a.u.}$ leads to a vanishingly small exponent, making the collisional ionization absolutely negligible.

%%%%%%%%%%%%%%%%%%%%%%%%%%%%%%%%%%%%%%
\section*{Acknowledgements}
We are happy to thank Thomas Gallagher, Johannes Nipper, Cenap Ates, Thomas Pohl, Hanna Schempp, Christoph Hofmann and Frank Grossmann for fruitful discussions. 
A.E. acknowledges financial support from the DFG under contract No. Ei 872/1-1.

\end{document}